\documentstyle[preprint,prb,aps]{revtex}
\begin{document}
\draft
\title{Solution of the Boltzmann equation \\
in a random magnetic field}
\author{Per Hedeg\aa rd\cite{berkeley} and Anders Smith,}
\address{\O rsted Laboratory, Niels Bohr Institute, Universitetsparken 5\\
DK-2100 Copenhagen \O, Denmark}
\date{\today}
\maketitle
\begin{abstract}
A general framework for solving the Boltzmann equation for a 2-dimensional
electron gas (2DEG) in random magnetic fields is presented, when the
random fields are included in the driving force. The formalism is
applied to some recent experiments, and a possible extension to
composite fermions at $\nu=1/2$ is discussed.
\end{abstract}

\section{Introduction}
Considerable interest both theoretically and
experimentally has recently focussed on the response of a 2DEG in spatially
inhomogeneous external fields. Examples include the so-called
Weiss-oscillations when a periodic electric field modulation is imposed%
,\cite{weiss} similar effects are predicted for periodic magnetic fields%
,\cite{kinesere} and random magnetic fields have been studied recently by
several authors.\cite{geim,geimetal,smithetal} In this paper we show
how to set up
a calculation of the transport coefficients of a 2DEG in a random magnetic
field. The magnetic fields in the experiments cited above are all fairly
low, well below the onset of Shubnikov-de Haas oscillations,
leading us to expect a semiclassical treatment to be valid. We explicitly
disregard any weak localization effects, which are well understood and can
be subtracted from the experimental results.

The plan of the paper is as follows: In
section \ref{prelimsec} we
describe the basic physical system and our model. In section \ref{bolt} we
discuss how to include the random magnetic field in the Boltzmann equation.
Section \ref{eigensec} and \ref{ressec}
are devoted to the calculation of the resistivity tensor. Finally, in
section \ref{applsec} we give
some applications of our theory to recent experiments, and discuss
suggestions for future experiments. We also briefly consider the
applicability of our theory to the Quantum Hall state at $\nu=1/2$.
The conclusion is in section \ref{concsec}.

\section{Preliminaries} \label{prelimsec}
Let us consider a 2DEG in a constant (external) electric field $\bbox{E}_0$,
and a spatially varying magnetic field applied perpendicular to the plane of
the electron gas: $\bbox{B}(\bbox{r})=(0,0,B(\bbox{r}))$
($\bbox{r}$ is in the plane).
There have been different realisations of the magnetic field: One,
originally proposed by Rammer and Shelankov\cite{rammer} for studying
weak localization effects in inhomogeneous magnetic fields, consists in
a type II superconducting gate on top of the heterostructure containing
the 2DEG. When above the lower critical field $B_{c1}$ for the superconductor,
an applied field will penetrate in flux tubes, each containing an integral
number of (superconductivity) flux quanta $\Phi_0^{\text{SC}}=h/2e$.
Because of disorder
in the superconductor the flux tube distribution will not be the Abrikosov
lattice, but a more or less random configuration, giving rise to a
randomly modulated field in the 2DEG. Another approach was taken by
Smith et al.\cite{smithetal} who deposited small lead grains on the surface
of a high mobility GaAs/AlGaAs heterostructure. For the grain sizes considered
Pb is a type I superconductor. Below $B_c$ there will be (partial)
flux expulsion from the grains, creating an inhomogenous magnetic field in the
2DEG. Still another possibility
is to deposit small grains of a permanent magnetic material on top of the
2DEG. The different kinds of inhomogeneity will give rise to
different field-dependences of the magnetoresistance.

We will model the random magnetic field by the following expression:
\begin{equation}
B(\bbox{r})=\sum_i b(\bbox{r}-\bbox{r}_i),
\end{equation}
where $\bbox{r}_i$ is the position of the $i$th impurity (flux tube,
lead grain,...) and $b$ the field produced by the impurity (including any
constant external field). We will assume
that the positions are randomly distributed over the sample and denote an
average over the positions $\bbox{r}_i$ by an overline.
Defining $B_0=\overline{B(\bbox{r})}$ we can write
\begin{equation}
B(\bbox{r})=B_0+\delta B(\bbox{r}),
\end{equation}
where $\delta B$ has zero average.

\section{The Boltzmann Equation} \label{bolt}
The standard derivation of the Boltzmann equation from the full
quantum mechanical transport equation relies on a separation of length
scales:\cite{transport} The external fields vary slowly in comparison
with $1/k_F$, while the rapidly varying impurity potentials are
included in a self-energy, giving rise to the collision term in the
Boltzmann equation. A previous attempt\cite{khaetskii} to include
random magnetic
fields in the Boltzmann equation has viewed the magnetic `impurities'
as (asymmetric) scatterers, i.\ e.\ has included them in the
scattering term on the right hand side. This, of course, can always be
done, even
when the correlation length $a$ of the modulation
(the London length or the size of the lead grains) is much greater
than $1/k_F$. However, when (as is the case in the experiments)
$a \gg 1/k_F$, the modulation $\delta B$ can just as well be treated
as an ordinary
external field in the driving force term of the Boltzmann equation.
(By contrast it would be
inconsistent to put $\delta B$ on the left hand side when $a\sim 1/k_F$.)
In the present paper we will take the latter point of view, i.\ e.\ put the
random magnetic field on the left hand side of the Boltzmann equation.
For the ordinary impurity scattering we will assume the relaxation time
approximation with a constant scattering time $\tau$.

The Boltzmann equation for the
distribution function $f(\bbox{r},\bbox{v})$ is then
\begin{equation}
\bbox{v}\cdot\frac{\partial f}{\partial\bbox{r}}+\bbox{a}(\bbox{r})\cdot
\frac{\partial f}{\partial \bbox{v}} = \left(\frac{\partial f}{\partial t}%
\right)_{\text{coll}},
\end{equation}
where the Lorentz force $m\bbox{a}(\bbox{r})$ is
\begin{equation}
\bbox{a}(\bbox{r}) = -\frac{e}{m}(\bbox{E}_0+ \bbox{v}\times\bbox{B(r)}),
\label{boltz}
\end{equation}
and
\begin{equation}
\left(\frac{\partial f}{\partial t}\right)_{\text{coll}} =
-\frac{f-f^0}{\tau}.
\end{equation}
We calculate to linear order in the external electric field $\bbox{E}$,
i.\ e.\ replace $-(e/m)\bbox{E}\cdot\frac{\partial f}{\partial \bbox{v}}
\rightarrow -e\bbox{E}\cdot\bbox{v}\frac{\partial f^0}{\partial E}$,
where $f^0$ is the Fermi-Dirac distribution. Introducing polar co-ordinates
$v,\phi$ for the velocity, at $T=0$ (the only case
considered here) $v$ only enters through
$\frac{\partial f^0}{\partial E}=-\delta(E-E_F)=-(m v_F)^{-1}%
\delta(v-v_F)$ and can be put equal to $v_F$. Writing
$f(\bbox{r},v,\phi)=f^0+g(\bbox{r},\phi)\delta(v-v_F)$ we get the following
equation for $g$:
\begin{equation}
\left\{ v_F \left( \begin{array}{c} \cos\phi \\ \sin\phi \end{array}
\right)\cdot
\frac{\partial}{\partial\bbox{r}} + (\omega_0+\delta\omega(\bbox{r}))
\frac{\partial}{\partial\phi}
+\frac{1}{\tau}\right\}g(\bbox{r},\phi) = -\frac{e}{m}\bbox{E}\cdot
\left( \begin{array}{c} \cos\phi \\ \sin\phi \end{array}
\right) .
\label{lign}
\end{equation}
Here $\omega_0=eB_0/m$ and $\delta\omega(\bbox{r})=e\,\delta B/m$. With
the following definitions (the $i$'s are for convenience):
\begin{eqnarray}
D_0 & = & i\left\{ v_F\left( \begin{array}{c} \cos\phi \\ \sin\phi \end{array}
\right)\cdot
\frac{\partial}{\partial\bbox{r}} + \omega_0
\frac{\partial}{\partial\phi}
+\frac{1}{\tau}\right\} \\
W & = & i \delta\omega(\bbox{r})\frac{\partial}{\partial\phi} \\
\chi(\phi) & = & - i \frac{e}{m}\bbox{E}\cdot
\left( \begin{array}{c} \cos\phi \\ \sin\phi \end{array}
\right),
\end{eqnarray}
we can write Eq. (\ref{lign}) as an operator equation
\begin{equation}
Dg \equiv (D_0 + W)g = \chi.
\end{equation}

The strategy is now to find the Green's function $G$ for $D$. Then we get
\begin{equation}
g(\bbox{r},\phi ) =
\int d\bbox{r}'\,d\phi'\, G(\bbox{r},\phi;\bbox{r}',\phi') \chi(\phi');
\label{g}
\end{equation}
and the current density
\begin{eqnarray}
\bbox{j} &=& -2 e \int\frac{d\bbox{p}}{(2\pi\hbar)^2}\bbox{v}
g(\bbox{r},\phi ) \delta(v-v_F) \nonumber \\
 & = & -\frac{1}{\pi}ne\int_0^{2\pi} d\phi \left(
\begin{array}{c}
\cos\phi\\ \sin\phi
\end{array} \right)
g(\bbox{r},\phi), \label{j}
\end{eqnarray}
where $n$ is the electron density.

\section{Eigenfunctions and Green's function for $D_0$} \label{eigensec}
It is straightforward to see that $D_0$ has the following complete set
of eigenfunctions and corresponding eigenvalues:
\begin{eqnarray}
\psi_{n\bbox{k}}(\bbox{r},\phi)&=&\frac{1}{\sqrt{2\pi
A}}e^{i\left[\bbox{k}\cdot(\bbox{r}-\bbox{R}(\phi))
- n\phi\right]}  \label{eigenf} \\
\lambda_n & = & n\omega_0 + \frac{i}{\tau}.
\end{eqnarray}
Here $A$ is the area of the sample and we have defined
\begin{equation}
\bbox{R}(\phi) = r_c \left( \begin{array}{c} \sin \phi \\ -\cos \phi
\end{array} \right),
\end{equation}
where $r_c=v_F/\omega_0$ is the average cyclotron radius.

The Green's
function is then given by
\begin{eqnarray}
G_0(\bbox{r},\phi;\bbox{r}',\phi') & = & \sum_{n \bbox{k}}
\frac{\psi_{n \bbox{k}}(\bbox{r},\phi)\psi_{n \bbox{k}}^*(\bbox{r}',\phi')}%
{n\omega_0 + i/\tau} \label{g0} \\
 & = & -\frac{i}{\omega_0}\frac{1}{e^{2\pi/\omega_0\tau}-1}
e^{[\phi'-\phi]/\omega_0\tau}\delta(\bbox{r}-\bbox{r}'-\bbox{R}(\phi)+
\bbox{R}(\phi')) \label{g0fin}
\end{eqnarray}
(see appendix \ref{appg0} for details).
We use the notation $[\theta]$ to denote
$\theta\pmod{2\pi}$. Note that $G_0$ is only a function of
$\bbox{r}-\bbox{r}'$ because of the translational invariance of $D_0$.
Inserting Eq.\ (\ref{g0fin}) in Eqs.\ (\ref{g}) and (\ref{j}) for the
current, the ordinary Drude formula is recovered.

\section{Calculation of the resistivity} \label{ressec}
We first have to consider how to perform the average over the random
magnetic field. Since the distribution function $g$ is the physically relevant
quantity, it is $g$ we have to average. From Eq.\ (\ref{g}) we see that
this means we have to find the averaged Green's function ($\chi$, of course,
only depends on the external electric field) for $D$. To do this we consider
the expansion of $D^{-1}$:
\begin{equation}
D^{-1}=(D_0+W)^{-1}=D_0^{-1} - D_0^{-1}WD_0^{-1} + D_0^{-1}WD_0^{-1}WD_0^{-1}-
\cdots . \label{d0w}
\end{equation}
We now average Eq.\ (\ref{d0w}) term by term. This is formally similar to
the quantum mechanical treatment of ordinary impurity scattering. Like in
quantum mechanics we can organize the terms into diagrams as shown in
figure \ref{diagram}. Summing the appropriate geometrical series we can
rewrite the equation for the averaged Green's function in terms of a
self-energy:
\begin{equation}
\overline{D^{-1}} = (D_0 + \Sigma)^{-1},
\end{equation}
where $\Sigma$ is the sum of all irreducible diagrams, i.\ e.\ diagrams
that do not fall apart when an internal line is cut,
as shown in figure \ref{selfen}.

We shall only go to second order in $W$, i.\ e.\ only keep the first diagram
in figure \ref{selfen}. The higher order diagrams (including the maximally
crossed ones, of which the first is shown in the figure) can in principle be
included; they are all finite and we do not expect them to change our
conclusions (as we have explicitly verified for the second diagram in
figure \ref{selfen}).\cite{unpub} There are no divergences corresponding to
`weak localization' in this purely semiclassical calculation.

We thus truncate the self-energy and write
\begin{equation}
\overline{D^{-1}}=(D_0- \overline{WD_0^{-1}W})^{-1}, \label{trunc}
\end{equation}
The `self-energy' $\overline{WD_0^{-1}W}$ is calculated as follows.
Using that
$W$ is Hermitian we get:
\begin{eqnarray}
\langle n\bbox{k}|\overline{WD_0^{-1}W}|n'\bbox{k}'\rangle & = &
\int d\bbox{r}\,d\phi\,d\bbox{r}'\,d\phi'\,
\overline{(W\psi_{n \bbox{k}}(\bbox{r},\phi))^*
G_0(\bbox{r},\phi;\bbox{r}',\phi')W\psi_{n' \bbox{k}'}(\bbox{r}',\phi')}
\nonumber \\
& = & \int d\bbox{r}\,d\phi\,d\bbox{r}'\,d\phi'\,
\overline{\delta\omega(\bbox{r})
\delta\omega(\bbox{r}')} \nonumber \\
& & \hspace{-3cm} \times[n+\bbox{k}\cdot\bbox{r}(\phi)]
[n'+\bbox{k}'\cdot\bbox{r}(\phi')]
\psi_{n \bbox{k}}(\bbox{r},\phi)^*\psi_{n' \bbox{k}'}(\bbox{r}',\phi')
G_0(\bbox{r},\phi;\bbox{r}',\phi'). \label{self}
\end{eqnarray}
Here $\bbox{r}(\phi)=r_c(\cos\phi,\sin\phi)$.
Now the averaging gives (see appendix \ref{appaver})
$\overline{\delta\omega(\bbox{r})
\delta\omega(\bbox{r}')}=f(\bbox{r}-\bbox{r}')$, where the correlation
function $f$ (which actually only depends on $|\bbox{r}-\bbox{r}'|$) depends
on the nature of the random magnetic field modulation. The averaging restores
translational invariance, meaning that the self-energy (\ref{self}) (and
therefore $D^{-1}$) is
diagonal in $\bbox{k}$. This is also seen explicitly by changing integration
variables to $\bbox{r}_1=\bbox{r}-\bbox{r}'$ and $\bbox{r}_2=\bbox{r}+
\bbox{r}'$. The only $\bbox{r}_2$-dependence of the integrand is then
through the factor $\exp(-i(\bbox{k}-\bbox{k}')\cdot\bbox{r}_2/2)$, giving a
$\delta_{\bbox{k},\bbox{k}'}$ upon integration. Therefore $G$ only depends
on $\bbox{r}-\bbox{r}'$. From Eq.\ (\ref{g}) we see that only ($A$ times)
the $\bbox{k}=
0$\ component is needed for the current. We can therefore put $\bbox{k}=0$ in
the self-energy (\ref{self}). Inserting the expression (\ref{g0fin}) for
$G_0$ we can use the delta function to do the $\bbox{r}_1$-integration,
giving
\begin{eqnarray}
\lefteqn{\langle n\bbox{0}|\overline{WD_0^{-1}W}|n'\bbox{0}\rangle}
\nonumber\\
& & = -\frac{i}{\omega_0}\frac{nn'}{e^{2\pi/\omega_0\tau}-1}
\frac{1}{2\pi}\int_0^{2\pi}d\phi\int_0^{2\pi}d\phi'
f(\bbox{R}(\phi)-\bbox{R}(\phi'))e^{[\phi'-\phi]/\omega_0\tau}e^{in\phi}
e^{-in'\phi'}.
\end{eqnarray}
The integrand is periodic in $\phi'$, so we can shift the limits on the
$\phi'$-integration to $\phi$ and $\phi+2\pi$. Changing variables to
$\theta=\phi'-\phi$ and using
\begin{equation}
|\bbox{R}(\phi)-\bbox{R}(\phi')|=2r_c|\sin\frac{\phi'-\phi}{2}|,
\end{equation}
the $\phi$-integration gives a
$\delta_{n,n'}$ (which was obvious from
the start: when $\bbox{k}=0$ the matrix element is also rotationally
invariant.)
The final result for the self-energy is
\begin{equation}
\Sigma_n\equiv \langle n\bbox{0}|\overline{WD_0^{-1}W}|n\bbox{0}\rangle =
-\frac{i}{\omega_0}\frac{1}{e^{2\pi/\omega_0\tau}-1}n^2
\int_0^{2\pi}f(2r_c\sin\frac{\theta}{2})e^{\theta/\omega_0\tau}e^{-in\theta}d\theta.
\label{selffinal}
\end{equation}
Now the left hand side of Eq.\ (\ref{trunc}) is diagonal and we get
\begin{equation}
\langle n\bbox{0}|\overline{D_0^{-1}}|n'\bbox{0}\rangle =
\frac{\delta_{n,n'}}{n\omega_0+i/\tau-\Sigma_n}.
\end{equation}
For the Green's function we get
\begin{equation}
\int d\bbox{r}G(\bbox{r},\phi;\bbox{r}',\phi')=\frac{1}{2\pi}\sum_n
\frac{e^{-in(\phi-\phi')}}{n\omega_0+i/\tau-\Sigma_n}.
\end{equation}
It is straightforward to do the remaining angular integrals in
Eq.\ (\ref{g}) and
(\ref{j}) to get the current density.
The result for the resistivity tensor is
\begin{equation}
\bbox{\rho} = \frac{m}{ne^2\tilde{\tau}}
 \left( \begin{array}{cc} 1 & \tilde{\omega}\tilde{\tau} \\
           -\tilde{\omega}\tilde{\tau} & 1 \end{array}\right),
\end{equation}
where we have used $\Sigma_{-n}=-\Sigma_{n}^*$ and
defined the renormalised quantities
\begin{eqnarray}
\tilde{\omega} & = & \omega_0 - \mbox{Re}\, \Sigma_1 \\
\frac{1}{\tilde{\tau}} & = & \frac{1}{\tau} - \mbox{Im}\, \Sigma_1.
\end{eqnarray}
We see that the change in $\bbox{\rho}$ is directly related to $\Sigma_1$:
\begin{eqnarray}
\frac{\Delta\rho_{xx}}{\rho_{xx0}} & = & -\tau \mbox{Im}\, \Sigma_1
\label{rhoxx} \\
\frac{\Delta\rho_{xy}}{\rho_{xy0}} & = & -\omega_0^{-1} \mbox{Re}\, \Sigma_1.
\end{eqnarray}

\section{Applications} \label{applsec}
In ref.\ \onlinecite{smithetal} we have applied the above theory to the
case where the field modulation is caused by the
Meissner flux expulsion from deposited grains of Pb. Furthermore we
also showed that the theory could be applied to the experiments by
Geim et al.\cite{geimetal} where
the field modulation was caused by flux tubes. This establishes that the
increase in the resistivity even in this case has an essentially
classical origin.

The general analysis of the magnetoresistance can be quite complex owing to
the three characteristic length scales in the problem:
the mean free path $l_{\tau}=v_F\tau$, the
magnetic field correlation length $l$, and the average cyclotron radius
$r_c=v_F/\omega_0$ (when the magnetic field is such that quantum effects
become important, a fourth length scale, $1/k_F$, enters).
In this section we shall  confine ourselves to considering the following
correlation function for different parameter values:
\begin{equation}
f(r)=n_i \delta\omega^2 \frac{\pi l^2}{2}e^{-r^2/2 l^2}, \label{frozcorr}
\end{equation}
which arises from an assumed gaussian field from a single impurity:
\begin{equation}
\delta\omega(r) = \delta\omega\, e^{-r^2/l^2}.
\end{equation}
This correlation function is applicable to the experiment
in ref.\ \onlinecite{smithetal}, when the modulation was dominated by
frozen flux in the grains (in this case $\delta\omega$
would depend on the external
field and go to zero as
$B_0\rightarrow B_c$), or the case when the modulation is caused by
permanent magnets.

We can find an explicit result for the change in the longitudinal
resistance for $B_0=0$ by expanding the integral in the self-energy
(\ref{selffinal}) (only $\theta\sim 2\pi$ contributes). We get
\begin{equation}
\frac{\Delta\rho_{xx}(B_0=0)}{\rho_{xx0}}=(\delta\omega\,\tau)^2
n_i \frac{\pi l^2}{2}
\int_0^{\infty} e^{-\left(\frac{l_{\tau}}{l}\right)^2 s^2 -s}ds.
\end{equation}
The overall shape of the magnetoresistance is determined
by the ratio $x=l_{\tau}/l$. When $x\lesssim 2$
there is a maximum in $\rho_{xx}$
at $B_0=0$, i.\ e.\ a negative magnetoresistance. As $x$ is increased $B_0=0$
is still a local maximum, but the magnetoresistance develops an intermediate
minimum and has a maximum for $B_0 > 0$. The location of the maximum is
approximately given by the condition that the average cyclotron radius should
equal $l$. When $x \gtrsim$\  4, $B_0=0$ becomes a local {\em minimum}.
In fig.\ \ref{plot1} we show a plot of the relative change of the
longitudinal resistance for $x$ varying between 1 and 100.
Fig.\ \ref{plot2} shows the relative change of the Hall resistance for
the same range of $x$'es. By varying
the electron density, the mobility,  and the size $l$ of the impurities,
the different
types of behaviour might be seen experimentally.
%Let us first investigate the case when $r_c, l \ll
%l_{\tau}$. In fig.\ \ref{expfit} we have shown the resulting
%longitudinal magnetoresistance (calculated from Eq.\ (\ref{rhoxx})).
%We see that as $l$ increases we change from a situation with a large
%negative magnetoresistance around $B_0=0$ to a situation with a
%maximum in $\rho_{xx}$ at finite $B_0$. The position of the maximum is
%given by the condition that the average cyclotron radius
%$v_F/\omega_0$ should equal $l$. By varying the density of the 2DEG,
%thereby varying $v_F$, this should be experimentally observable.
We
note furthermore that it is possible to have a very large enhancement of the
zero-field resistance in a random magnetic field (for reasonable
parameters we can easily get a 20-fold increase).

The problem of the (quantum mechanical) localization of a particle in a
random magnetic field has received a lot of attention.\cite{locmag} As we
have seen we can get a negative magnetoresistance around $B_0=0$,
reminiscent of weak localization. This enhancement of the resistance
could be interpreted as a classical
sign of `localization' in the random magnetic field. However, varying the
parameters it is also possible to get a local minimum around $B_0=0$, so
there is probably no simple connection to the quantum mechanical theory
of localization (although we always get an {\em enhancement} of
the resistance).

We end this section by some speculations regarding the possibility of
applying our theory to the Quantum Hall $\nu=1/2$-state. As shown in
the seminal paper by Halperin et al.\cite{halperin} it is possible to
make a Chern-Simons gauge transformation and transform the
$\nu=1/2$-state to a system of fermions (socalled composite fermions,
which can be viewed as an electron with two flux-tubes attached to it)
in zero average magnetic field. The composite fermions interact
through a fictitious Chern-Simons magnetic field given by
\begin{equation}
\delta b_{\text{CS}} = 2 \Phi_0\, \delta\rho,
\end{equation}
where $\Phi_0=h/e$ is the flux quantum and $\delta\rho$ is the
deviation of the electron density from the average density $n$. In a
static approximation $\delta\rho$ is nonzero simply because of the
electrostatic coupling to the impurities. We then have a system of
fermions moving in a static, random magnetic field, i.\ e.\ the system
considered above. If the impurities are fully ionized we would have
$n_i=n$, i.\ e.\ a much higher impurity density than in the grain
experiments, and $l$ is typically much smaller than before (of the order of
500 \AA). In our model this will give rise to a broad
minimum in $\rho_{xx}$ around $B_0=0$ (that is around $B_{1/2}$) for
our simple model correlation function (see fig.\ \ref{plot3}).

In fact, we are here dealing with the limit $x\gg 1$, in which case
the self-energy (\ref{selffinal}) can be evaluated analytically for
moderate fields. We get
\begin{equation}
\label{largex}
\frac{\Delta \rho_{xx}}{\rho_{xx0}} = \left(\frac{\pi}{2}\right)^{\frac{3}{2}}
n_i l^2 \frac{(\delta\omega\tau)^2}{x}
\coth\left(\frac{\pi}{\omega_0\tau}\right)
\end{equation}

This is consistent with the behaviour
seen experimentally.\cite{jiang,du,kang,leadley} We also note that typical
values of $\rho_{\nu=1/2}/\rho_0$ are in the range 25--500.
To do quantitative comparisons
with experiment there is an additional complication:
Since the impurities give rise
both to the magnetic field and the impurity scattering it is
inconsistent to put them on different sides of the Boltzmann equation,
when in reality they are strongly correlated. Instead the potential
scattering should be brought to the LHS of the Boltzmann equation as a
random electric field. This does not contradict the discussion in the
beginning of section \ref{bolt}, since $k_F d \sim 15 \gg 1$ in a
typical QHE sample.\cite{halperin} ($d \sim$\ the distance from the
2DEG to the impurity layer, sets the distance over which the impurity
potential varies appreciably).
Calculations are currently underway, addressing
this issue.

\section{Conclusion} \label{concsec}
We have shown how to calculate the magnetoresistance of a 2DEG in a random
magnetic field in the semiclassical approximation. The magnetic field was
included in the driving force term in the Boltzmann equation. Furthermore
we have applied our results to different experiments and shown how the
different types of random magnetic fields gives rise to the differences in
the magnetoresistance. Finally we speculated on the possible relevance
of our theory to the $\nu=1/2$ Quantum Hall state.

We would like to thank Mads Nielsen and Kasper Juel Eriksen for discussions.

\appendix
\section{Green's function for $D_0$}
\label{appg0}
Inserting the expression (\ref{eigenf}) for the eigenfunctions into Eq.\
(\ref{g0}) we get
\begin{equation}
G_0(\bbox{r},\phi;\bbox{r}',\phi') =
\frac{1}{2\pi}\sum_n\frac{e^{in(\phi'-\phi)}}{n\omega_0+i/\tau}
\frac{1}{A}\sum_{\bbox{k}}e^{i\bbox{k}\cdot(\bbox{r}-\bbox{r}'
-\bbox{R}(\phi)+\bbox{R}(\phi'))}.
\end{equation}
The $\bbox{k}$-sum just gives a delta function. The sum over $n$ involves
calculating
\begin{eqnarray}
h(\phi) & = & \frac{1}{2\pi}\sum_n\frac{e^{in\phi}}{n+i\alpha} \nonumber \\
& = & \frac{1}{2\pi}\sum_{m=-\infty}^{\infty}\int_{-\infty}^{\infty}
e^{i(2\pi m+\phi)x}\frac{1}{x+i\alpha}dx,
\end{eqnarray}
where we have used the Poisson summation formula. The integrand has a
simple pole in $x=-i\alpha$. If $2\pi m+\phi>0$, we close the contour in the
upper
half plane and get 0. If $2\pi m+\phi<0$, we close in the lower and pick up a
contribution $-2\pi ie^{(2\pi m+\phi)\alpha}$ from the pole:
\begin{equation}
h(\phi) = -i\sum_{2\pi m< -\phi} e^{(2\pi m +\phi)\alpha}.
\end{equation}
Since $h$ is periodic in $\phi$ we can choose the argument to lie
between $0$ and $2\pi$ and get
\begin{eqnarray}
h(\phi) & = & -i e^{\alpha[\phi]}\sum_{m=-\infty}^{-1}e^{2\pi\alpha m}
\nonumber \\
& = & -i e^{\alpha[\phi]}\frac{e^{-2\pi\alpha}}{1-e^{-2\pi\alpha}}.
\end{eqnarray}
With $\alpha=1/\omega_0\tau$ we then get Eq.\ (\ref{g0fin}).

\section{Calculation of the correlation function}
\label{appaver}
We have to calculate $\overline{\delta\omega(\bbox{r})
\delta\omega(\bbox{r}')}=(e^2/m^2)\overline{\delta B(\bbox{r})
\delta B(\bbox{r}')}$. The averaging consists of integrating over the $N$
positions (assumed independent) of the impurities. We get
\begin{eqnarray}
\overline{\delta B(\bbox{r})
\delta B(\bbox{r}')} & = & \overline{B(\bbox{r})B(\bbox{r}')} -B_0^2
\nonumber\\
& = & \sum_{i,j}\overline{b(\bbox{r}-\bbox{r}_i)b(\bbox{r}'-\bbox{r}_j)} -B_0^2
\nonumber\\
& = & \frac{1}{A^N}\sum_{i,j}\int\prod_{k=1}^N d\bbox{r}_k \,
b(\bbox{r}-\bbox{r}_i)b(\bbox{r}'-\bbox{r}_j) -B_0^2.
\end{eqnarray}
The $N(N-1)$ terms with $i\neq j$ are equal to
$A^{N-2}(\int d\bbox{R}\, b(\bbox{R}))^2
=A^N B_0^2/N^2$ (since $B_0=\overline{B}=N\overline{b}$), and the $N$ terms
with
$i=j$ are equal to
$A^{N-1}\int d\bbox{R}\, b(\bbox{r}-\bbox{R})b(\bbox{r}'-\bbox{R})$. Dropping
the
term of order $1/N$ we get
\begin{equation}
\overline{\delta B(\bbox{r})
\delta B(\bbox{r}')} = n_i \int d\bbox{R}\,
b(\bbox{r}-\bbox{R})b(\bbox{r}'-\bbox{R}),
\end{equation}
where $n_i=N/A$ is the density of impurities. The correlation function thus
becomes
\begin{equation}
f(\bbox{r})=n_i \int d\bbox{R}\, b(\bbox{r}+\bbox{R})b(\bbox{R}).
\label{appf}
\end{equation}
It is only a function of $r$. Note that it does not matter if we write
$\delta b$ for $b$ in (\ref{appf}), as $b$ and $\delta b$ only differ by a
constant of order $1/N$.

\begin{figure}
\caption{The perturbation series for the full propagator (Green's function)
$G$. The unperturbed propagator $G_0$ is denoted by a thin arrow. A cross
denotes $W$, and the dashed lines an impurity average.}
\label{diagram}
\end{figure}
\begin{figure}
\caption{The (irreducible) self-energy $\Sigma$. The calculations in the text
approximates $\Sigma$ by the first diagram in the series.}
\label{selfen}
\end{figure}
\begin{figure}
\caption{Relative change in longitudinal
resistance of a 2DEG in a
random magnetic field with
the correlation function (\protect\ref{frozcorr}). Parameters as
follows: $n=4 \ 10^{15} \ \text{m}^{-2}$, $n_i= 10^{9} \
\text{m}^{-2}$, $db=0.03 \ \text{T}$. The curves show (from the top) $x=
1,2,3,5,10,20,100$. We are varying $l$ and $l_\tau$ such that $l l_\tau=
l_0^2$, where $l_0=8.1 \ \mu\text{m}$ is constant, to keep the curves on the
same scale.}
\label{plot1}
\end{figure}
\begin{figure}
\caption{Relative change in Hall resistance for $x$ varying between 1 and 100,
with parameters as in
fig.\ \protect\ref{plot1}. $x=1$ corresponds to the lowest curve.}
\label{plot2}
\end{figure}
\begin{figure}
\caption{The relative change in the longitudinal resistance with parameters
as follows: $n=8 \ 10^{15} \ \text{m}^{-2}$, $n_i=8 \ 10^{15} \
\text{m}^{-2}$, $l=500 \ \text{\AA}$, $\mu = 100 \
\text{m}^2/\text{Vs}$, $db=0.03 \ \text{T}$. The change in the Hall
resistance (not shown) is very small, of the order of $10^{-3}$.}
\label{plot3}
\end{figure}
\end{document}